\documentclass{emulateapj}
\usepackage{epsfig}
\usepackage{graphicx}
\usepackage{url}
\usepackage[dvips]{color}

\newcommand{\sgr}{\mbox{SGR\,J0501+4516~}} 
\newcommand{\sgrnos}{\mbox{SGR\,J0501+4516}}
 
\newcommand{\sgrbnos}{\mbox{SGR\,J0418+5729}}

\shorttitle{Photospheric radius expansion magnetar bursts}
\shortauthors{Watts et al.}

\begin{document}

\title{Photospheric radius expansion during magnetar bursts}
\author{Anna L. Watts\altaffilmark{1,2}, Chryssa
  Kouveliotou\altaffilmark{3}, Alexander J. van der Horst\altaffilmark{4}, Ersin G{\"o}{\u g}{\"u}{\c s}\altaffilmark{5},
  Yuki Kaneko\altaffilmark{5}, Michiel van der Klis\altaffilmark{1}, Ralph
  A.M.J. Wijers\altaffilmark{1}, Alice K. Harding\altaffilmark{6} and Matthew G. Baring\altaffilmark{7}}
\altaffiltext{1}{Astronomical Institute ``Anton Pannekoek'', University of
  Amsterdam, Postbus 94249, 1090 GE Amsterdam, the Netherlands}
\altaffiltext{2}{Email: A.L.Watts@uva.nl}
\altaffiltext{3}{Space Science Office, VP62, NASA Marshall Space
  Flight Center, Huntsville,
  AL 35812, USA}
\altaffiltext{4}{NASA Postdoctoral Program Fellow, NASA Marshall Space
  Flight Center, Huntsville, AL 35805, USA}
\altaffiltext{5}{Sabanc\i~University, Orhanl\i-Tuzla, \.Istanbul
  34956, Turkey}
\altaffiltext{6}{NASA Goddard Space Flight Center, Greenbelt, MD
  20771, USA}
\altaffiltext{7}{Department of Physics and Astronomy, MS-108, Rice University, P. O. Box 1892, Houston, TX 77251-1892, USA}

\begin{abstract}
On August 24th 2008 the new magnetar SGR 0501+4516
(discovered by {\it SWIFT}) emitted a bright
burst with a pronounced double-peak structure in hard X-rays,
reminiscent of the double-peak temporal structure seen in some bright thermonuclear bursts on accreting
neutron stars.  In the latter case
this is due to Photospheric Radius Expansion (PRE): when the flux
reaches the Eddington limit, the photosphere expands and cools so that
emission becomes softer and drops temporarily out
of the X-ray band, re-appearing as the photosphere settles back down.  We consider the factors necessary to
generate double-peaked PRE events, and show that such a mechanism
could plausibly operate in magnetar bursts {\it despite the vastly
  different emission process}.  Identification
of the magnetic Eddington limit in a magnetar would constrain magnetic field
and distance and could, in principle, enable a measurement of
gravitational redshift.  It would also locate the emitting region at the
neutron star surface, constraining the burst trigger mechanism.
Conclusive confirmation of PRE events will require more detailed radiative
models for bursts.  However for
SGR 0501+4516 the predicted critical flux (using the
magnetic field strength inferred from timing and the distance
suggested by its probable location in the Perseus arm of our Galaxy) is consistent
with that observed in the August 24th burst.  
\end{abstract}

\keywords{pulsars: individual (SGR 0501+4516) — stars: magnetic fields — stars: neutron — X-rays: bursts}

\maketitle

\section{Introduction}

Type I X-ray bursts are thermonuclear explosions caused by unstable
burning of light elements in the surface layers of accreting neutron
stars (for a review see e.g. \citealt{Lewin93}).    Luminosities frequently reach the Eddington limit, at which
point radiation pressure lifts the surface layers from the star in a
Photospheric Radius Expansion (PRE) episode.  One of the hallmarks of
bright PRE bursts, when observed with sufficient time resolution, is a pronounced double peak structure in the
X-ray lightcurve \citep{Hoffman78, Hoffman80, Paczynski83,
  Ebisuzaki84, Lewin84, Tawara84,
  Vacca86, Haberl87}.  As the photosphere moves outwards the temperature
drops and the energy of the emitted photons falls below the X-ray
band, leading to an apparent drop in countrate. As the photosphere
contracts again the temperature rises, and there is a second brighter peak in
X-ray emission \citep{Paczynski83}.  If one looks at the bolometric rather
than the X-ray lightcurve, the double-peak structure almost completely
disappears.

PRE bursts have proven extremely useful in studies of accreting
neutron stars since they act as standard candles, yielding 
distance \citep{vanParadijs78, vanParadijs81, Kuulkers03, Galloway03, Galloway06a}.  
PRE events can also be used to place constraints on mass and radius,
and hence the dense matter equation of state (see for example
\citealt{Damen90}, \citealt{Galloway08} and \citealt{Ozel09}).  

On August 24th 2008 the newly-discovered magnetar SGR 0501+4516
\citep{Barthelmy08} emitted a bright burst with a pronounced 
double-peaked structure (Figure \ref{fig:gbmlcs}).  This motivated us to consider whether multi-peak PRE
events might be possible in magnetar bursts, despite the vastly
different emission mechanism (magnetic rather than thermonuclear).  The existence of a magnetically modified Eddington limit has
been discussed in the literature for a number of years \citep{Paczynski92,Ulmer94,Thompson95,Miller95,Israel08b}.  However the
prospect of PRE and multi-peaked burst lightcurves has never been considered before, perhaps in part due
to the considerable
uncertainty that still exists over the magnetar burst trigger and
emission mechanism.  

We start in Section \ref{t1bursts} with the thermonuclear burst case,
and identify the factors essential to the generation of double-peaked
PRE bursts.  In Section \ref{magflares} we move onto magnetar bursts,
and consider
whether there are burst emission scenarios where these 
conditions might be met.  We conclude that PRE events might plausibly occur
in magnetar bursts under certain circumstances.  In Section
\ref{consequences} we consider what could
be learnt from an unambiguous identification of a PRE magnetar burst.  In
addition to constraining stellar properties (as for the X-ray burst case) it
would also constrain the burst trigger and emission mechanism.  

In Section \ref{aug24} we return to the event that motivated our study
and ask whether the 
bright burst from SGR 0501+4516 on August 24th 2008 could be an example of
PRE.  We show that the predicted critical flux, using the
magnetic field strength inferred from timing and the distance
suggested by its probable location in the Perseus arm of our Galaxy, is consistent
with that observed.  Confirmation, however,
requires the development of more detailed radiative models for the
bursts.  In Section \ref{discussion} we broaden our scope to include
the other magnetars, and assess whether we could or should have seen
PRE episodes from these sources. The magnetar population includes
objects classifed as Soft Gamma Repeaters (SGRs, which tend to burst
frequently) and Anomalous X-ray Pulsars (AXPs, which burst
less frequently).  We cover both SGRs and AXPs in our discussion (see
\citealt{Woods06} a comprehensive review of the properties of the
different types of magnetars).   We conclude in Section \ref{conclusions}.

\section{PRE in thermonuclear X-ray bursts}
\label{t1bursts}

For thermonuclear bursts to exhibit PRE and for the X-ray lightcurve
to be double-peaked, four basic conditions must be met:  

\newcounter{itemcounter}
\begin{list}{\textrm{\arabic{itemcounter}.}}{\usecounter{itemcounter}\leftmargin=1em}
\item Flux has to be emitted from an optically thick region.
\item There must be a critical luminosity where radiation pressure
    can balance gravitational and other confining forces on the
    emitting material.
\item{Opacity must increase with radius.}
\item{The emitting region must cool as the photosphere expands.}
\end{list}
  
\subsection{Condition 1:  Location of emitting region}
\label{bcond1}

In order for there to be a photosphere, the initial emission has to
occur in an optically thick region and propagate outwards.  For X-ray
bursts this condition is easily met, since the thermonuclear runaway
that triggers the burst can only occur at the
base of the neutron star ocean, where density $\sim 10^6$ g/cm$^3$
(see \citealt{Bildsten98} and references therein).  In this regime optical depth is much
greater than unity and a photosphere will exist.    

\subsection{Condition 2:  The existence of a critical luminosity}
\label{bcond2}

The critical, or `Eddington' luminosity can be calculated by considering the balance
between radiation pressure and gravity in the emitting matter.  For
accreting neutron stars we can neglect other confining forces since
magnetic fields are weak.  If the
accreting material is fully ionized, then radiation exerts a force
primarily on the electrons via Thomson scattering.  Coulomb attraction
between protons and electrons means that radiation must act
against a gravitational force set predominantly by the nuclear mass.
Under these conditions, force balance in Newtonian gravity yields

\begin{equation}
 L_\mathrm{Edd} = \frac{4 \pi cGM}{\kappa} 
\label{eddnewtgen2}
\end{equation}
where $M$ is the gravitational mass of the neutron star (see
\citealt{Lewin93} and references therein). For non-magnetic systems subject only to Thomson scattering,
the opacity $\kappa$ is defined as

\begin{equation}
   \kappa =  \frac{\sigma_T n_e}{\rho}
 \label{opac}
\end{equation} 
where $\sigma_T$ is the Thomson cross-section, $n_e$ the number density of
electrons and $\rho$ the density.  The precise value of $\kappa$
depends on the composition of the accreted material
due to contributions in the X-ray band from bound-free transitions for
heavier elements.  As we shall see in Section \ref{magflares}, the intense magnetic field in 
magnetars introduces profound modifications
to the determination of $\kappa$.  For thermonuclear bursting sources,
the inferred magnetic fields are too low to have such an effect.

When considering observable quantities, we need to take into account
how General Relativity modifies this expression.  The gravitational
force is stronger by a factor $(1+z)$ where

\begin{equation}
1+z =\left(1 - \frac{2GM}{Rc^2}\right)^{-1/2}
\end{equation}
and $R$ is the radial distance from the center of the neutron star as measured
by a local observer \citep{Lewin93}.  This
modifies the force balance equation, and means that the critical luminosity
as measured by a local observer at the
photosphere\footnote{Calculations and observations show that the outflow
  velocities are much smaller than the speed of light
  \citep{Ebisuzaki83, Kato83, Paczynski86b, Joss87, intZand10}.  This
  means that the difference between the luminosity measured by an
  observer moving with the photosphere, and by a stationary observer
  at the same radial distance, can be neglected.} 

\begin{equation}
L_\mathrm{cr} = \left(\frac{4\pi c GM}{\kappa}\right)(1+z)
\label{eddphot}
\end{equation}
For an observer at infinity, 

\begin{equation}
L_{\infty} = L_\mathrm{ph}(1 + z_\mathrm{ph})^{-2}
\label{eddinf}
\end{equation}
where the subscript `ph' refers to the photosphere.  This means that for a
distant observer, 

\begin{equation}
L_{\mathrm{Edd}, \infty}= \frac{4\pi c GM}{\kappa
    (1+z_\mathrm{ph})}
\label{eddfinal}
\end{equation}
\citep{Lewin93}.  For typical neutron star parameters ($M =
1.4M_\odot$, $R_\star = 10$ km) and solar to He-rich composition for the
accreting material, this yields
{$L_{\mathrm{Edd}, \infty} \approx (2-3.7) \times 10^{38}$ ergs/s at
  touchdown (when $R = R_\star$).  This value
is comparable to the luminosities observed during the brightest Type
I X-ray bursts \citep{Kuulkers03, Galloway08b}.

\subsection{Condition 3: Increase in opacity with radius}
\label{bcond3}

The local critical luminosity falls as radius increases, as $(1+z)$
(Equation \ref{eddphot}).  However the luminosity of the propagating
photons also falls, as $(1+z)^2$
(Equation \ref{eddinf}).  These General Relativistic effects
impose further conditions on the occurrence of PRE bursts.  Consider what would happen if the opacity were
constant in the atmosphere.  In this case, the ratio of the luminosity to
the critical luminosity would vary with radius as

\begin{equation}
\frac{L}{L_\mathrm{cr}} \propto (1+z)
\end{equation}
reducing outwards \citep{Paczynski86a}. The luminosity
emerging from the photosphere would not reach the critical value unless
the luminosity at depth exceeded the local limit.  This could only be achieved via
convection, which would largely prevent envelope (and hence
photospheric radius) expansion \citep{Paczynski86a,Woosley04,Weinberg06}.
 
This problem can only be bypassed if the critical luminosity 
is higher at greater depths within the photosphere.  
It turns out that this is possible, because
opacity depends on temperature $T$ \citep{Hanawa82}
when Klein-Nishina modifications become significant.  
The opacity should more properly be
written as: 

\begin{equation}
\kappa = \frac{\kappa_0}{1 + (2.2\times 10^{-9}~T)^{0.86}} 
\label{opactemp}
\end{equation}
\citep{Paczynski83}, where $\kappa_0$ is the non-magnetic Thomson opacity given in Equation
(\ref{opac}).  The burning layer is hotter ($> 10^8$ K even at
ignition, up to  $\sim 10^9$ K during the burst) than the photosphere
($\sim 10^7$ K), so the opacity is much lower at greater depths, i.e. lower altitudes.  This permits a high flux to
propagate out of the hot burning layer without a large convective zone
being present.  

The increase in opacity with altitude as the plasma cools 
is also essential for sustained photospheric radius expansion to occur. 
If this were not
the case then once the photosphere
started to expand and cool (Section \ref{bcond4}), the ratio
$L/L_\mathrm{cr}$ would quickly drop below 
unity, halting expansion. We note in passing that in the magnetar application
(Section \ref{magflares}), the gradients of the magnetic field define a stronger 
dependence of opacity and associated critical luminosity on altitude
than the gravitational redshift influences identified here.

\subsection{Condition 4: Cooling of emission region with expansion}
\label{bcond4}

To obtain a double-peaked X-ray lightcurve, the photosphere must cool
as it expands in response to radiation pressure.  For X-ray bursts
this is thought to occur because the emission is quasi-blackbody, and
the luminosity at the photosphere remains close to critical.  For blackbody emission 

\begin{equation}
L_{\infty} = 4\pi (R_{\infty})^2 \sigma (T_{\mathrm{eff}, \infty})^4
\end{equation}  
where $\sigma$ is the Stefan-Boltzmann constant, $R_\infty = R (1+z_\mathrm{ph})$ is the photospheric radius as measured by the distant
observer, and $T_\mathrm{eff}$ is the effective temperature
\citep{Rybicki04}.  If photospheric luminosity remains at the
critical value given by Equation (\ref{eddphot}), as suggested by
theoretical simulations \citep{Paczynski83,Kato83, Ebisuzaki83,
  Ebisuzaki84, Quinn85, Paczynski86a, Paczynski86b, Joss87, Nobili94,
  Weinberg06}, then we obtain

\begin{equation}
T_{\mathrm{eff}, \infty} =  \left[\frac{c GM}{\sigma \kappa}
\right]^{1/4} R^{-1/2} (1 + z)^{-3/4}
\end{equation}
The redshift factor $(1 + z)^{-3/4}$ increases as the photosphere
expands, but more slowly than the $R^{-1/2}$ factor, so the observed
temperature $T_{\mathrm{eff}, \infty}$ falls as $R$ increases.

How well does this simple model hold up for real X-ray bursts?  X-ray
bursts are indeed generally well-fit by a blackbody spectrum
\citep{Swank77, Lewin93,Galloway08b}, although there are some
deviations \citep{Damen89, Damen90, Damen90b, vanParadijs90, Kuulkers03}.  Spectral fitting to multi-peak
PRE bursts supports the picture of temperature falling as radius
expands, although again there are some minor discrepancies in the
observations.  Bolometric luminosity, for example, often continues
to rise all the way through until touchdown at the surface, 
rather than rising and
then falling, as predicted by Equation (\ref{eddfinal})
\citep{Galloway08b}.  There has yet to
be any serious systematic effort to quantify and resolve the remaining
discrepancies:  however non-Planckian spectra \citep{vanParadijs82}, compositional effects \citep{Galloway06a},
obscuration or scattering by the accretion
flow \citep{Damen90b, Galloway08}, and clearing of the inner parts of the disk
by the expanding
photosphere \citep{Shaposhnikov03} may all play a role.

\section{PRE in magnetar bursts}
\label{magflares}

Magnetar burst fluences follow a power law distribution (log$N$-log$S$) with an index of $\sim-1.7$ (see \citealt{Woods06};
also \citealt{Gogus01}). The bulk of this emission is in weak, soft,
events that occur in bunches during a burst-active episode. At times
several hundreds of these bursts have been recorded during a 24 hour
period of magnetar activity. Occasionally, sources emit much brighter
events (intermediate bursts) and very rarely, giant flares - only
three of the latter have ever been recorded. Small burst luminosities
vary between $(10^{-2} - 10^3$) $L_{\rm Edd}$, while the giant flare
luminosities can reach up to $10^7 L_{\rm Edd}$, where $L_{\rm Edd}$ is
the non-magnetic Eddington limit.  To determine whether any of these
apparently super-Eddington bursts could in principle exhibit
multi-peak behaviour due to PRE 
episodes, we must assess whether the four conditions
necessary for the occurrence of this phenomenon, identified in Section
\ref{t1bursts}, can be met by magnetars.  It is possible, of course,
that additional conditions may have to be met for PRE to operate
during magnetar bursts:  however it seems reasonable to start with the
four conditions that we know are required for PRE to occur in
thermonuclear X-ray bursts.  

The magnetar problem differs from the X-ray burst problem in several
key respects.  The burst mechanism is magnetic rather than thermonuclear, but there is
as yet no agreement on the trigger mechanism or emission site
location.  The strong magnetic field alters many of the emission
properties \citep{Harding06}:  in particular, scattering depends on
polarization, with E-mode (electric field vector polarized
perpendicular to the magnetic field) scattering suppressed
compared to the O-mode (electric field vector parallel to the magnetic
field) unless photons stream along field lines. The situation will also change
depending on whether we are discussing emission from open or closed
field line regions, since radiative transport across magnetic
field lines is strongly inhibited relative to that along fields in neutron 
star magnetospheres. We will examine the 
various scenarios that are currently envisaged, and assess
whether there are any circumstances under which
multi-peak PRE behaviour might be possible.  

\subsection{Condition 1:  Location of emitting region}
\label{magcond1}

There is consensus that the underlying cause of the
bursting activity (as 
well as many other magnetar properties) is the decay of the strong
magnetic field \citep{Woods06}.  This results in the field twisting
into a configuration that eventually becomes unstable
\citep{Braithwaite06a, Braithwaite06b}.  The bursts are generated by
rapid rearrangement of the magnetic field, and the 
formation and dissipation of localized currents.  External
reconfiguration is likely to involve reconnection (due to the many
instabilities that can operate in a plasma) although this may not
occur in all bursts \citep{Duncan04}.  What is not clear yet is
exactly where the dissipation and emission occurs: in the crust, at the surface of
the star, high in the magnetosphere, or a combination of all three?  
For a well-defined photosphere to exist emission must come from an
optically thick region, so emission in an optically thin region would not be compatible with the multi-peak PRE
hypothesis.  

Part of the uncertainty over emission region stems from the fact that
the trigger  
for magnetar bursts is still not known. For there to be sporadic
bursting activity, there has to be some 
barrier to magnetic reconfiguration that yields on the rise timescale
of the bursts.  \citet{Thompson95}
argue that the gate is the solid crust, which is placed under stress
by the untwisting core magnetic field.  In this picture the crust
fails when the twist exceeds its yielding strain, and the
rise time is set by the shear-wave crossing time of the crust.  As the
crust slips the exterior field moves along with it, leading to
reconfiguration and possible reconnection.  In this case it seems
likely that initial energy release would occur either in the
crust or just above, in the zones of maximum magnetic
shear.  

In the alternative trigger model of \citet{Lyutikov03} the crust moves
plastically in 
response to the magnetic stresses \citep{Jones03}, with a twist
building up in the current-carrying magnetosphere.  Eventually the
magnetosphere loses equilibrium and a
tearing instability in the plasma triggers reconnection.  The 
timescale in this case is set by the growth time of the resistive
tearing mode.  The required rise time would seem to
place the initial emission locale in low density regions high in the
magnetosphere \citep{Duncan04}: emission close to the surface
would lead to rise times slower than those observed. 

The trigger mechanisms described above operate on the rise timescales
of the bursts ($\sim$ 0.01s).  To explain the durations of the bursts
($\sim 0.1$s for single-peaked bursts, \citealt{Gogus01}), there must
in addition be some way of prolonging 
the emission.  Either the initial trigger must recur (perhaps with
one event sparking the next) or there
has to be some way to store the initial energy and release it over a
longer timescale.   

In the crust failure model, a recurrent trigger would require a
network of crust ruptures, with slip along one `fault' setting
off the next.  The activity would have to be confined to a
relatively small part of the stellar surface, since if the rupture
spread rapidly to the far side of the star then the slow rate of rotation
would ensure that any emission close to the surface would disappear
from view.  The
possibility of avalanches of reconnection is discussed by 
\citet{Lyutikov03} in the context of the magnetospheric trigger
model. In this picture emission probably does not take place in an optically
thick region unless precipitating particles generated by the avalanche
impact the surface.  The heated surface would then radiate on a thermal
timescale. 

The alternative is some kind of storage mechanism that leads to
emission over a longer period.  Two possibilities have been
discussed in the literature: crust vibrations, and the formation of a
trapped pair-plasma 
fireball.  In the crust vibration model the initial impulse excites torsional
oscillations of the
crust \citep{Duncan98}. This is known to occur in the rare and highly
energetic giant flares
\citep{Israel05, Strohmayer05, Strohmayer06, Watts06}, although
vibrations have not yet been detected after the regular bursts.  The
excitation of crust vibrations is certainly plausible if the trigger
is crust failure:  whether vibrations
could be excited to a 
significant degree by magnetospheric reconnection alone is less
clear.  The oscillating crust couples (via the charged lattice) to the
field lines, generating Alfv\'en waves which lead to particle
acceleration and prolonged emission.  Where in the magnetosphere this
excitation and emission might occur would depend on the amplitude of the
oscillations, with 
larger motions coupling to longer field lines.  If the motions are
 strong enough, however, the energy available may be
sufficient to generate an optically thick pair plasma. For very high
release rates this may lead to trapped 
fireball formation (see below) but for slower injection rates the
plasma can form an optically thick corona in which there is a steady balance between
injection rate and radiation rate.  This effect has been invoked to
explain the initial smooth tail in 
the giant flare lightcurves \citep{Thompson01a}, and  \citet{Gogus01}
have discussed its possible role in the smaller bursts.  

The other possibility is the formation of a magnetically trapped,
optically thick fireball that gradually leaks  
radiation.  \citet{Thompson95} argued that such a phenomenon was an
inevitable consequence of very rapid energy generation in a closed
field line region.  This could be due to either 
reconnection \citep{Thompson95} or the development of a Quantum
Electrodynamic (QED) shock
\citep{Heyl05a}.  A fireball could
therefore be 
formed from either of the two trigger mechanisms provided that local energy
generation rate is high enough and within a closed field line region.
The rapid generation of Alfv\'en waves or relativistic particles leads
to the formation of a dense optically-thick thermal plasma of
e$^\pm$ pairs and 
$\gamma$-rays.  The charged pairs effectively cannot cross the magnetic
field lines, and their density (and hence scattering opacity) is
sufficiently high that they trap radiation.  The fireball cools and
contracts due to radiative diffusion from a thin surface layer, with the bulk
of the radiation leakage occurring close to the stellar surface where scattering
is suppressed (see Section \ref{magcond2}).  The opacity here will be dominated by electrons
and ions ablated from the neutron star surface (especially if the
emergent flux
is close to the magnetic Eddington limit) which form the photosphere.  The heated surface
exposed as the fireball 
retreats will also continue to emit radiation as it cools. 

The fireball model has been very successful at explaining the later
decaying tail phase of giant flares \citep{Thompson01a}.   Spectral
fitting indicates that the emitting area falls while the temperature of the radiation remains
roughly constant at the level expected for
the photosphere of a trapped fireball in a magnetic field in excess of
$B_\mathrm{QED} = 4.4\times 10^{13}$ G \citep{Thompson95, Feroci01}.  In the early stages
(the `smooth tail') the photosphere is dominated by pairs (see above), while in
the later stages it reverts to one dominated by electrons and ions
ablated from the stellar surface. Whether fireballs form in the
smaller bursts is still not clear (energy release may not occur at a
fast enough rate, \citealt{Gogus01}), although the spectra are similar to
those of the decaying tails of the giant flares \citep{Woods06}.

To summarize, however,
there are viable scenarios for both optically thick and 
optically thin emission. In the magnetospheric
instability picture optically thick emission is much less likely.
Within the crust slippage model emission may occur from optically thick
regions, particularly if (a) reconnection occurs in or close to the
surface layers, or (b) a trapped fireball or pair corona forms.  In
this case the formation of a 
well-defined photosphere is possible. In fact, as we will see in
Section \ref{magcond2}, the two different polarization modes will have
 spatially distinct photospheres due to their different scattering
properties.

\subsection{Condition 2: The existence of a critical luminosity}
\label{magcond2}

Scattering opacities in a magnetar strength field are strongly
modified compared to the non-magnetic case outlined in Section
(\ref{bcond2}).  In closed field line regions, the magnetic field can
also provide an additional non-negligible confining force to balance
radiation pressure.  Both of these factors increase the Eddington
limit.  

\subsubsection{Reduction in scattering}
\label{magcond2a}

\citet{Paczynski92} was the first to address the apparently
super-Eddington luminosities emitted during SGR bursts.  He considered
the case of energy release deep within the surface layers, with
photons diffusing out at a rate limited by 
electron conductivity or photon opacity.   In a magnetic field $B$ both Thomson and Compton
cross-sections are reduced for photon energies $E_\gamma = \hbar\omega$ lower than
the electron cyclotron energy $E_c$, where

\begin{equation}
E_c = \hbar\omega_c = \mathrm{11.58~keV}\left(\frac{B}{10^{12} ~\mathrm{G}}\right),
\end{equation}
because electrons cannot easily
move perpendicular to the magnetic field.  Consider the case where $\left(E_\gamma/E_c\right)^2 \ll 1$, 
a condition that is generally met for magnetar bursts in
the energy bands that we are considering; the cyclotron energy $E_c$ for the 
fundamental Compton scattering resonance falls in the X-ray band only at
altitudes of $\gtrsim 20$ stellar radii for magnetars.  For the O-mode the
scattering cross-section is

\begin{equation}
\sigma_\parallel/\sigma_T \approx \sin^2\theta +
\left(E_\gamma/E_c\right)^2 \cos^2\theta,
\label{scatpar}
\end{equation}
where $\theta$ is the 
angle between the direction of propagation of the photons and the
magnetic field.  For the E-mode,

\begin{equation}
\sigma_\perp/\sigma_T \approx  \left(E_\gamma/E_c\right)^2. 
\label{scatperp}
\end{equation}
These magnetic Thomson domain results can be deduced from
Eq. (16) of \citet{Herold79}.  Full QED numerical evaluations of the
polarization-averaged  
magnetic Compton cross section are displayed for different $\theta$ in
Fig.~3 of  
\citet{Herold79} and Fig.~3 of \citet{Daugherty86}.

Scattering is therefore suppressed for both polarizations for
radiation flowing along open field lines, and is always suppressed for
the E-mode.  This means that the E-mode and O-mode photospheres will
be spatially distinct whenever the polarization states are decoupled, with 
the E-mode photosphere extending deeper into the surface layers. 
If polarization mode-switching via scattering is prolific, then the two 
photospheric scales become coupled, a nuance that is addressed below.

A useful quantity for estimating the radiative flux at large optical depth 
is the Rosseland mean opacity} $\bar{\kappa}$, where

\begin{equation}
\frac{1}{\bar{\kappa}} = \left[\int_0^\infty \frac{1}{\kappa_\nu}
  \frac{\partial B_\nu (T)}{\partial T} d\nu \right]\bigg/ \left[\int_0^\infty 
  \frac{\partial B_\nu (T)}{\partial T} d\nu\right].
\end{equation}
In this expression, $\kappa_\nu$ is the monochromatic opacity
at the photon
frequency $\nu = \omega/2\pi$, and $T$ is the temperature.  In field-free 
regions, $\kappa_\nu$ can be represented by Equation (\ref{opac}) 
or Equation (\ref{opactemp}).
The function $B_\nu$ is the Planck function \citep{Rybicki04}. 
For the highly-anisotropic conditions imposed by the
strong magnetar fields, $\kappa_\nu$ represents a weighted average 
over photon angles with respect to the magnetic field.  However, it should be 
remarked that technically, the Rosseland mean opacity is most
conveniently employed for almost isotropic photon populations, i.e.
applied to radiative transfer problems in the interiors of normal stars.
Notwithstanding, it is still a useful measure, and here,
as expected, $\kappa_\nu$ is lowest
for photons streaming along field lines.  Note also that for X-ray bursts
(Section \ref{bcond2}) there is no need to use the Rosseland mean
opacity, because the 
scattering cross-section $\sigma_T$ does not depend on photon energy. 
Hereafter, $\bar{\kappa}$ can be interpreted as a 
photon polarization-dependent quantity, or as an average of photon 
polarizations, as needed.

Equation (\ref{eddnewtgen2}) for the critical luminosity now becomes

\begin{equation} 
L_\mathrm{crit} = \frac{4\pi c G M}{\bar{\kappa}}.
\end{equation}
This {\it magnetic Eddington} luminosity depends on magnetic field
strength and temperature as well as the stellar parameters. For photons
streaming along field lines it follows that

\begin{equation}
\frac{L_\mathrm{crit}}{L_\mathrm{Edd}} \approx
\left(\frac{\omega_\mathrm{c}}{\omega}\right)^2
\label{eddred}
\end{equation}
where $L_\mathrm{Edd}$ is as given in Equation
(\ref{eddnewtgen2}). Anisotropies and, as we shall see below,
mode-switching between polarizations, will profoundly influence this ratio.  

To compute $L_\mathrm{crit}$, \citet{Paczynski92} uses the fact that
$\omega = kT/\hbar$ and then assumes blackbody emission, so that the
critical flux $F_\mathrm{crit} = \sigma T^4$.  Under this assumption,
and using Equation (\ref{eddnewtgen2})
for $L_\mathrm{Edd}$, one can rewrite Equation (\ref{eddred}) as

\begin{equation}
\frac{L_\mathrm{crit}}{L_\mathrm{Edd}} \sim 2 \left(\frac{B}{10^{12}
    ~\mathrm{G}}\right)^{4/3} \left(\frac{g}{2\times 10^{14}
    \mathrm{~cm/s}^2}\right)^{-1/3}. 
\label{eddlimpacz}
\end{equation}
This is the estimate used for magnetic Eddington limit in, for example,
\citet{Israel08b}.  

Subsequent authors have re-visited this calculation and made a number
of corrections and additions.  \citet{Thompson95}, for example, noted
that Equation (\ref{scatperp}) is
only valid when plasma density is low \citep{Herold79}.  If plasma
density is higher (so that the plasma frequency approaches the cyclotron frequency),
as it might be for surface rather than
magnetospheric emission, the scattering cross-section is higher (see also
\citealt{Meszaros92} and \citealt{Miller95}).  For
emission from the neutron star surface, Equation (\ref{scatperp}) would become:

\begin{equation}
\sigma_\perp/\sigma_T \approx  \frac{1}{\sin^2\theta}
\left(E_\gamma/E_c\right)^2, 
\label{scatperpsurf}
\end{equation}
increasing the scattering cross-section of the E-mode.
\citet{Thompson95} re-compute the magnetic Eddington limit for this cross-section, again
assuming blackbody emission.  The coefficients that they find are slightly different to
those derived by \citet{Paczynski92}, but
to a factor of order unity Equation (\ref{eddred}) still applies.

\citet{Ulmer94} and \citet{Miller95} considered the important effect of
scattering between polarization states on the critical luminosity.
\citet{Miller95} demonstrated that the emergent luminosity is
dominated by E-mode photons, since O-mode photons above the E-mode
photosphere will continue to scatter into the E-mode and then escape
from the star.  The radiation force, however, is dominated by the
O-mode due to the much higher scattering cross-section.
\citet{Miller95} uses order of magnitude estimates of the
mode-scattering to show that the luminosity
  which eventually emerges in the O-mode, $L_\parallel$, is given by:

\begin{equation}
L_\parallel  \sim  0.1 \frac{\omega}{\omega_c} L_\mathrm{tot}
\end{equation}
where  $L_\mathrm{tot}$ is the total luminosity in both polarization
states.  For the small number of photons that end up in this state,
the scattering cross-section $\sim \sigma_T$ (Equation
\ref{scatpar}). For the atmosphere to remain hydrostatic, one
 requires $L_\parallel < L_\mathrm{Edd}$, so that

\begin{equation}
L_\mathrm{tot} \lesssim 10 (\omega_c/\omega) L_\mathrm{Edd}
\label{eddlimmiller}
\end{equation}
This estimate, which was then verified using
Monte Carlo simulations of radiative transfer \citep{Miller95}, is
lower than that 
obtained by \citet{Paczynski92} and \citet{Thompson95}.  

 The critical luminosity may in fact be lower
still due to other scattering processes that operate in
a magnetized neutron star atmosphere including vacuum polarization and mode
switching, the proton cyclotron resonance, and bound-free
absorption \citep{Miller95, Thompson02}.  Photon splitting will also
 increase the fraction of 
O-mode photons, further increasing the radiation force
\citep{Miller95, Thompson01a}.  One additional factor that none of the above calculations include is
the effect of gravitational redshift, something that is taken into account in all
of the estimates of critical flux for X-ray bursts (Section
\ref{bcond2}).   This should be included in the estimates
of observed critical luminosity
if the photosphere is close to the neutron star surface.  Such general relativistic corrections can be introduced by an 
effective blueshift to the photon frequency entering into
Equations~(\ref{eddred}) or (\ref{eddlimmiller})
that acts to reduce the critical luminosity.  However, the enhancement of the 
magnetic field strength in the local inertial frame (see e.g.
\citet{Gonthier94} for the dipolar case) 
partially offsets this reduction by effectively blueshifting the cyclotron frequency.

\subsubsection{Magnetic confinement  effects}
\label{magcond2b}

In Section \ref{magcond2a} we saw that the magnetic field reduces
scattering for radiation propagating both parallel to and across field
lines, increasing the critical flux over the non-magnetic limit
derived in Section \ref{bcond2}.  In this Section we will consider
the effect of 
magnetic confinement. The magnetic field resists the motion of charged
particles across field lines, thereby contributing an additional term
to the force balance equation and
 increasing the critical flux for closed field line regions.  

The field
necessary to confine the plasma can be estimated by requiring that
magnetic pressure exceed radiation pressure (assuming that radiation
pressure dominates gas pressure):

\begin{equation}
\frac{B^2}{8\pi} \gg \beta \frac{4\sigma T^4}{c}
\end{equation}
where $\beta$ depends on the angular distribution of the radiation
field:  it is 1/3 for isotropic radiation \citep{Lamb82}. Note that this estimate is only
valid for blackbody radiation, though it can be readily adapted to treat any luminosity per unit area 
passing through a surface element.  This simplifies \citep{Ulmer94, Miller95} to the
requirement that 

\begin{equation}
\left(\frac{B}{10^{12}~\mathrm{G}}\right) > \left(\frac{T}{170~\mathrm{keV}}\right)^2
\label{eq:magpress_crit}
\end{equation}
For magnetar bursts this condition is met provided the photospheric
radius is below about ten stellar radii.
Note that because magnetic
pressure acts perpendicular to the field, plasma can always move rapidly
along field lines.  In such cases, using a magnetohydrodynamic interpretation for the electromagnetic 
contribution to the stress-energy tensor, the left hand side of 
Equation~(\ref{eq:magpress_crit}) is replaced by a much smaller combination of 
the field and the plasma speed.  Hence, in
order to achieve confinement one therefore
needs closed field geometries, and even in this case matter will
 migrate towards the points where the field is weakest.  For a
dipole field this means towards the equator, and away from the stellar
surface. 

In the closed field line regions matter can be confined out to the point
where the pressure of free streaming photons exceeds the dipole
magnetic energy density.  At this radius, the optical depth will
drop to $\le 1$.  \citet{Thompson00} showed that this occurs for radii
greater than $R_A$ where

\begin{eqnarray}
\frac{R_A}{R_\star} \sim  280 \left(\frac{B_\star}{10
    B_\mathrm{QED}}\right)^{1/2} \left[ \left(\frac{10^{44}
    \mathrm{~ergs}}{E_\mathrm{burst}}\right)\left(\frac{\Delta
    t_\mathrm{burst}}{100\mathrm{~s}}\right)\right]^{1/4}
\end{eqnarray}
$E_\mathrm{burst}$ and $\Delta t_\mathrm{burst}$ are the energy and
duration of the burst respectively.  For typical SGR bursts, $R_A \gg R_\star$.  So in a closed field line
region, emitting plasma can be confined close to the 
stellar surface for luminosities far in
excess of the values derived in Section \ref{magcond2a} for open field
line regions.  

\subsection{Condition 3:  Increase in opacity with radius}
\label{magcond3}

In Section \ref{bcond3} we showed that GR effects would stifle PRE
unless opacity increased with altitude.  The same GR effects must apply
to magnetar bursts if the emission site is
close to the stellar surface.  For X-ray bursts the variation in
opacity with depth comes from the temperature dependence of the
opacity. For magnetar bursts the magnetic field dependence of the
opacity (Section \ref{magcond2a}) can provide a similar, albeit much
stronger, effect.

The magnetic field strength quickly falls off
with radius (for a dipole field as $1/R^3$), leading to a rapid
increase in opacity with height above the stellar
surface.  Evidence of this increase in scattering comes from the
strong rotational pulse profiles seen during the decaying tails of
lightcurves from the giant flares.   Radiation emitted from the base
of a trapped fireball (Section \ref{magcond1}) is thought to be collimated by
the increase in scattering opacity, forming highly focused jets of
X-ray emission \citep{Thompson95, Thompson01a}.

\subsection{Condition 4: Cooling of emission region with expansion}
\label{magcond4}

For thermonuclear bursts the expansion and cooling of the emitting
region follows very simply from the fact that the emission is, for the
most part, well
modelled by a blackbody (Section \ref{bcond4}).  
Demonstrating that the emitting region is expanding and cooling for
magnetar bursts is not as straightforward.
Although the majority of magnetar bursts are relatively soft (compared
to gamma-ray bursts, for example - although not to X-ray bursts, see \citealt{Woods06}), they
are not well 
fit by simple blackbody spectra\footnote{For an exception see \citet{Woods05}.}.  Multi-component spectral models containing
one or two blackbodies have had some 
success \citep{Feroci04, Olive04, Nakagawa07,Esposito07, Israel08b,
  Esposito08} but it is not yet clear to what extent these models are
physical rather than phenomenological.     

The failure of simple blackbody models is however not unexpected,
since radiative transfer effects in such strong magnetic fields should
substantially modify any initially thermal spectrum, of the type that
we might expect, for example from a trapped fireball
\citep{Thompson95,Thompson01a} or other optically thick region \citep{Paczynski92}.  Lower energy 
photons, for example, scatter less and can 
hence escape from deeper, hotter parts of the atmosphere.  The
radiation at low energies should thus exceed that expected for simple
blackbody emission \citep{Ulmer94, Lyubarsky02}.  Photon splitting and
merging will also be important in modifying the spectrum
\citep{Miller95, Thompson01a} at energies above around 30--50 keV.  

At present, modelling of magnetar burst emission, and the atmospheric
response of a magnetar to a 
flux at or exceeding the magnetic Eddington limit, is not sufficiently
advanced to permit us to make firm predictions for spectral evolution (see \citealt{Harding06} for an extended discussion of the
difficulties inherent in modelling radiative transfer for magnetar
bursts, which include the vastly disparate mean free paths
  for the two polarization modes).  
We are therefore not yet in a position to say conclusively
how, if PRE does occur and an underlying thermal region expands and cools,
this would be reflected in the emergent spectrum.  It seems logical,
however, that we should expect at least a drop in the overall energy
of emergent photons.

\section{Consequences of identifying PRE from a magnetar}
\label{consequences}

Detailed modelling of the type done for PRE X-ray bursts has not been
done for magnetar bursts, so it is not possible to say conclusively
whether the mechanism would work on the observed
timescales\footnote{Theoretical and observational studies for X-ray bursts indicate that the photosphere
  can expand at speeds of up to 0.01c, so large-scale expansion and
  contraction on timescales of 0.1s is not unreasonable
  \citep{Ebisuzaki83, Kato83, Paczynski86b, Joss87, intZand10}.}.  However
it certainly seems that it is plausible.  The
four conditions necessary for multipeak PRE bursts (Section
\ref{t1bursts}) can be met within some of the 
envisaged emission scenarios (Section \ref{magflares}), 
particularly those involving radiation from an optically thick pair
corona or trapped fireball into an open field line region.   In this
Section we will consider what could be learnt if we were able to
identify a PRE episode during a magnetar burst.

\subsection{Burst emission mechanism}

As outlined in Section \ref{magcond1} the mechanism responsible for
magnetar bursts is still not known.  Two possible trigger mechanisms
have been identified:  crust rupturing \citep{Thompson95} and
explosive magnetic reconnection \citep{Lyutikov03}.  Both are capable
of generating rise times that match the observations.  In addition
there needs to be some way of generating prolonged emission.  In the
crust fracture model, magnetic reconnection \citep{Thompson95} or QED
instabilities \citep{Heyl05a} generate an optically thick e$^\pm$
plasma fireball.  However, as outlined in \citet{Harding06} this model
does not explain all 
aspects of the spectra, although it does explain the durations of the
bursts by delaying energy release.  What could generate the prolonged
emission in the reconnection model is not entirely clear. A trapped fireball may be formed;
alternatively the duration is related to
the timescale over which repeated accelerations take place.  

An identification of PRE would confirm that
the emission was taking place in an optically thick region, hence
ruling out purely magnetospheric emission mechanisms.  The mechanism
identified by \citet{Lyutikov03} could still be responsible for the
initial burst trigger but would have to be augmented by some optically
thick radiation storage mechanism.  It would also confirm that
emission is occurring via open field line regions (since the critical flux
for closed field line regions would be much higher than the fluxes
seen during most normal bursts).

\subsection{Stellar properties}

The critical luminosity depends on magnetic field strength and the
gravitational field of the star (Section \ref{magcond2}).  The flux at which a PRE episode
occurs can therefore be used to confirm estimates of source distance
and magnetic field strength obtained via other means (supernova
remnant/globular cluster association and spin-down rate, for
example).  

Given trusted values of the source distance and magnetic field, one would then be
able to use the critical flux to measure the gravitational redshift of
the neutron star.  With high enough time resolution one should be able
to track changes in redshift as the photosphere lifts off, expands, and then touches back
down.  To obtain a constraint on mass and radius, one would need to
measure redshift at take-off or touchdown.  However a measurement at
maximum expansion would (in theory) tell you how far off the surface
the photosphere had risen.   Measuring magnetar redshifts
would be extremely interesting, since in X-ray burst sources 
 such estimates can be contaminated by the presence of the
accretion disk \citep{Galloway08}.

\section{The 2008 August 24 burst from \sgrnos}
\label{aug24}
 
\sgr was discovered with {\it Swift} when it became active on 2008 August 22 \citep{Barthelmy08,Enoto09,Rea09}. 
The source emitted a series of very intense bursts during the next 13
days, triggering GBM 26 times \citep{Fishman08,Kouveliotou08}. 
Subsequent {\it RXTE} observations revealed a period and a period derivative,
enabling an estimate of the average dipole magnetic field of $1.9\times10^{14}$ G, 
placing the source among the magnetar candidates \citep{Woods08,Israel08a}. 
This is the first magnetar candidate seen from the Galactic anticenter direction, 
suggesting a location in the Perseus arm of our Galaxy at a distance of $1.95\pm 0.04$~kpc \citep{Xu06}. 
If this distance is confirmed, \sgr would be one of the two closest magnetar candidates to Earth, 
together with \sgrbnos, discovered in 2009 \citep{vanderHorst09}, which also likely resides in the Perseus arm \citep{vanderHorst10}.
 
\begin{figure}
\begin{center}
\includegraphics[width=\columnwidth]{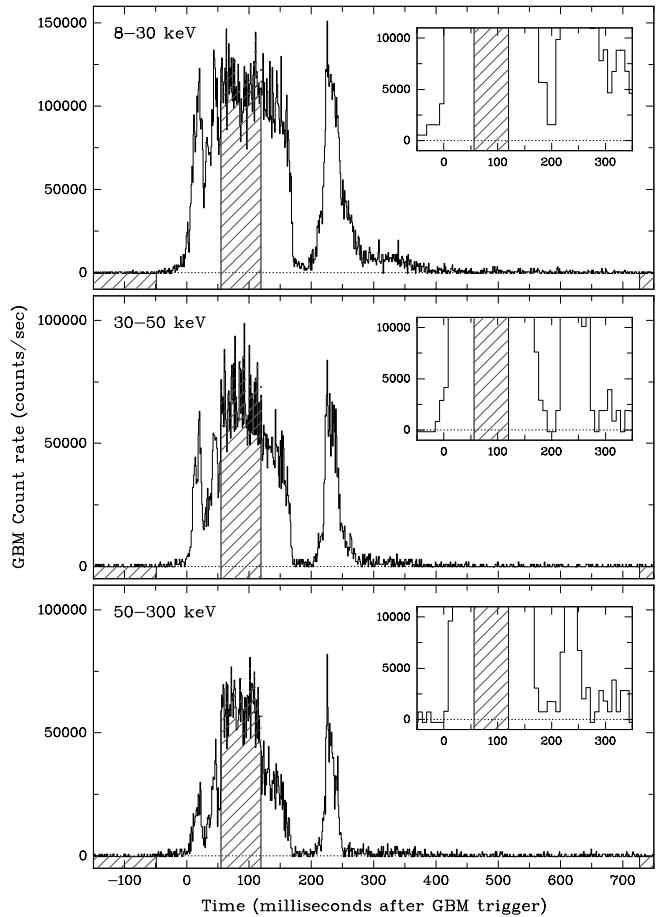}
\end{center}
\caption[]{Light curves of the bright burst on 2008 August 24 from \sgr in three energy ranges. 
The insets are a zoom-in to show the energy dependence of the gap size between the two bright pulses. 
The hashed area indicates the part of the light curve that was not used for spectral analysis (see also text).}
\label{fig:gbmlcs}
\end{figure}

On 2008 August 24, GBM recorded a very bright burst from \sgr with a pronounced double peak structure, 
and most importantly with the flux between peaks dropping almost to background levels (Figure \ref{fig:gbmlcs}). 
This peculiar burst light curve resembled those of X-ray bursts 
where PRE had been observed \citep{Lewin93}, 
motivating us to search for similar temporal and spectral signatures. 
One of the typical PRE X-ray burst characteristics is the drop of the flux to almost background level, 
after the initial pulse, followed by a second, more intense pulse. 
The gap between pulses appears above 6 keV in X-ray bursts and is longer for larger photon
energies. 
Although in the August 24 SGR burst the first pulse is actually more intense than the second, 
we show in Figure \ref{fig:gbmlcs} that there is an energy dependence of the gap size 
between pulses in the burst (see Figure \ref{fig:gbmlcs} insets).  The
gap (from 125 to 225 ms after GBM trigger) is much more pronounced at
higher energies.  

Spectrally, the burst was observable between $8-300$ keV and was so bright 
that it saturated the High Speed Science Data Bus of the GBM Data Processing Unit \citep{Meegan09}.  
As a result, part of the light curve is artificially cut off and cannot be used for any reliable analysis 
(see hashed area in Figure \ref{fig:gbmlcs}). 
We therefore performed spectral analysis outside the affected interval. 
We used the Time Tagged Event (TTE) data of NaI detectors 2 and 5 \citep{Meegan09},  
both with source angles to the detector normal of $\sim46^o$, binned at 8~ms resolution. 
We fitted the full spectral range of the NaI detectors ($8-1000$ keV), 
excluding a few energy channels around the Iodine K-edge at $\sim33$ keV, 
using the spectral analysis software package RMFIT\footnotemark{}. 
\footnotetext{R.S.~Mallozzi, R.D.~Preece, \& M.S.~Briggs, "RMFIT, A Lightcurve and Spectral Analysis Tool," 
\copyright 2008 Robert~D.~Preece, University of Alabama in Huntsville, 2008} 
We fitted various spectral models to the data: power law (PL), black body (BB), 
power law with an exponential cut off (Comptonized), optically-thin thermal bremsstrahlung (OTTB), 
and combinations of PL and BB, and two BB functions. 
The best fits for both the time-integrated and time-resolved analysis were obtained with the Comptonized function. 
The time-integrated spectrum\footnote{Computed using the non-hashed
  parts of the light curves in Figure \ref{fig:gbmlcs}:  times -0.049 to 0.055 and 0.119 to 0.727 seconds.} is best fit with an index of $-0.22\pm0.03$ and $E_{\rm{peak}}=36.3\pm0.2$ keV. 
The resulting photon and energy flux ($8-300$ keV) are $472\pm2$ ph/s/cm$^2$ and $(1.968\pm0.009)\times10^{-5}$ erg/s/cm$^2$, respectively.

\begin{figure*}
\begin{center}
\includegraphics[angle=-90,width=\textwidth]{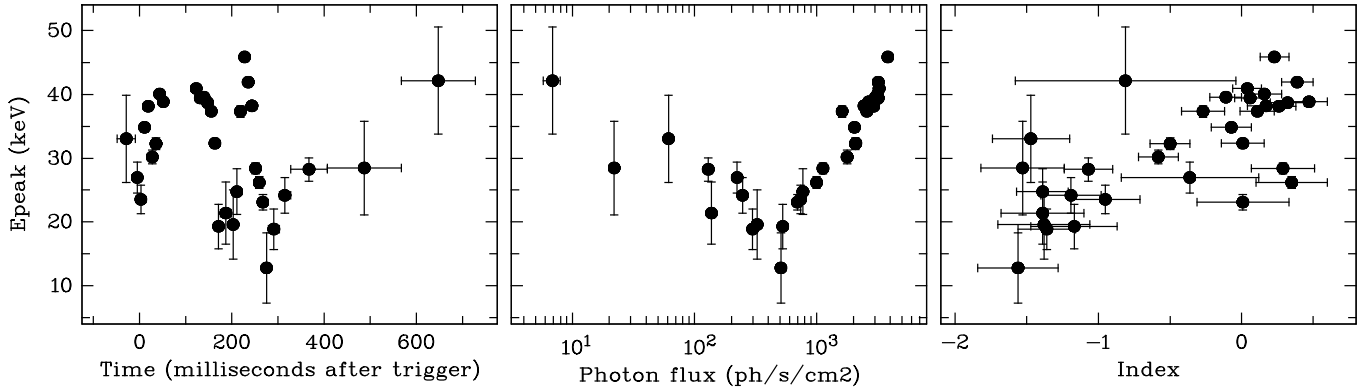}
\end{center}
\caption[]{Time-resolved spectral analysis of the bright 2008 August 24 burst from \sgrnos:
$E_{\rm{peak}}$ versus time (left panel), $E_{\rm{peak}}$ versus photon flux (middle panel),
and $E_{\rm{peak}}$ versus power-law index (right
panel). $E_{\rm{peak}}$ drops during the dip in the lightcurve between
125 and 225 ms after GBM trigger.}
\label{fig:epeak}
\end{figure*}

Our time-resolved analysis displays strong spectral evolution during the event, which is shown in Figure \ref{fig:epeak}. 
We find that the $E_{\rm{peak}}$ correlates strongly with the source flux when the burst  is bright (photon flux $>400$ ph/s/cm$^2$). 
This correlation breaks down at lower fluxes, where we even see an anti-correlation, 
in particular in the gap between the two bright pulses and in the tail  of the burst.  
Furthermore, there is a correlation between the Comptonized power-law index and $E_{\rm{peak}}$, 
with indices around $0$ for the highest and $\sim-1.5$ for the lowest $E_{\rm{peak}}$ values.

Since we were only able to analyze a part of the event, we cannot determine the peak or the total burst luminosity (or energy).
We can, however, put a lower limit on the peak flux, namely the one of the second pulse which has a lower peak flux than the saturated first pulse.
The lower limit on the 8-millisecond peak energy flux ($8-300$ keV) is
$(1.95 \pm 0.03)\times10^{-4}$ erg/s/cm$^2$.

Using Equation
(\ref{eddlimmiller}) we can estimate the critical
magnetic Eddington flux as

\begin{eqnarray}
F_\mathrm{crit} & \sim & 2\times 10^{-2} ~\mathrm{ergs/cm}^2\mathrm{/s}
\left(\frac{B}{10^{14} ~\mathrm{G}}\right) \left(\frac{1
  ~\mathrm{keV}}{E_\gamma} \right) {} \nonumber \\ & & \times
\left(\frac{1~\mathrm{kpc}}{d}\right)^2
\left(\frac{L_\mathrm{Edd}}{2\times 10^{38} ~\mathrm{ergs/s}}\right)
\label{elmillermod}
\end{eqnarray}

For $d = 1.95$ kpc and $B = 1.9\times 10^{14}$ G this yields $F_\mathrm{crit} \sim 3\times 10^{-4} $ ergs/cm$^2$/s
for photon energies of 36 keV (the $E_\mathrm{peak}$ value for the
time-integrated burst spectrum).  Inclusion of the gravitational
redshift would lower the critical flux by 25\%, to $\sim 2\times 10^{-4}$ ergs/cm$^2$/s. This is comparable to the lower limit on the peak flux we estimated for the August 24th event,
and also with the 2-millisecond peak flux of $2 \times 10^{-4} $
ergs/cm$^2$/s measured by Konus-Wind \citep{Golenetskii08}.  This
lends plausibility to the idea that we might be seeing an event that
reaches the magnetic Eddington limit and then undergoes PRE.  

\begin{figure}
\begin{center}
\includegraphics[width=\columnwidth]{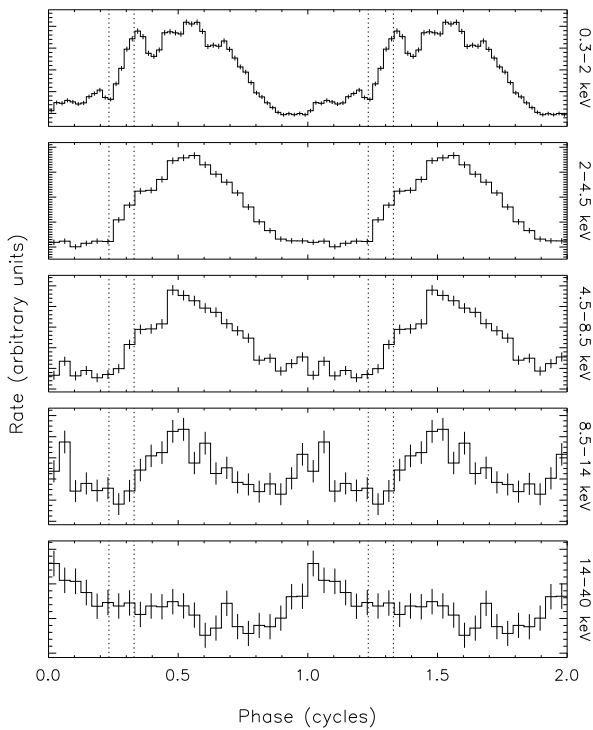}
\end{center}
\caption[]{Energy resolved pulse profiles (0.3-40 keV) generated using
  {\it RXTE} and {\it XMM-Newton} data from
  Aug 22-24 2008. The closest {\it RXTE} pointing (2-40 keV) ends about 3000 s
  before the burst and the one after starts about 5000 s later.  The
  {\it XMM-Newton} observations (0.3-2.0 keV) were taken a day before the burst.  At
  higher energies than those shown the pulse profile is consistent with random
  fluctuations.  The dashed lines indicate the phase interval within
  which the event took place.}
\label{fig:phasexte}
\end{figure}

We also investigated the light curve of the event versus the phase of
the spin of the source, to try to identify whether the burst happens
on an open or closed field line region (see below).  The burst data
were Earth barycentered and folded using the ephemeris 
obtained with {\it RXTE}.  Pulse profiles constructed from {\it RXTE}
and {\it XMM-Newton} data are shown in Figure \ref{fig:phasexte}.  Full
details of the spin and pulse profile analysis can be found in a
companion paper ({G{\"o}{\u g}{\"u}{\c s}} et al. in preparation). 

The burst happens during the rising part of the pulse profile below 10 keV, 
before the pulse maximum and before the pulse onset above 10 keV
(dashed lines, Figure \ref{fig:phasexte}). If
the model of \citet{Thompson95} is accurate, pulse maximum corresponds
to an open field line region, since the jets of
radiation that form the main pulses escape along open field lines.
This suggests that this particular burst occurs at the
same rotational phase as an open field line region.
This is consistent with the idea that you have to be on an open field line
region in order to get radius expansion at reasonable luminosities
(Section \ref{magcond2}).  

\section{Discussion}  
\label{discussion}
\subsection{Models for multi-peak magnetar bursts}

Multi-peaked magnetar bursts are a relatively common phenomenon.
These are usually assumed to be superpositions of smaller bursts,
due to the broad distribution of wait times between burst peaks
\citep{Gogus01}.  So is the 2008 August 24th burst from \sgr really anything special?  
It was certainly rather unusual compared to
most of these bursts.  Having the flux
dropping to near-zero at all is rare \citep{Woods06}. The profile -
particularly the very rapid drop in emission before the secondary rise
- is also odd.  Most multi-peaked bursts, by contrast, 
have a longer decay time than rise time.  It therefore seems
reasonable to postulate that a different mechanism might be
responsible for the double-peaked nature of this very bright burst.    

\subsection{Consequences for other magnetar bursts} 

If bursts that reach the magnetic Eddington limit are possible then we
should consider the consequences for other sources and bursts.  Table
\ref{sdata} shows the open field line magnetic Eddington limit predicted for other
magnetars, for all magnetars with estimates of distance and magnetic
field strength, for photon energies of 50 keV\footnote{Data taken from the McGill magnetar catalogue, http://www.physics.mcgill.ca/$\sim$pulsar/magnetar/main.html}.

\begin{deluxetable}{cccc}
\tabletypesize{\small}
\tablewidth{0pt}
\tablecaption{Estimated magnetic Eddington limits for known magnetars \label{sdata}}
\tablehead{\colhead{Source}  & \colhead{Distance} &
  \colhead{Magnetic} &
  \colhead{$F_\mathrm{crit}$\tablenotemark{$\dagger$}} \\ & (kpc)\tablenotemark{a} &
  field & (erg/cm$^{2}$/s) \\ &  &
  ($\times 10^{14}$G)\tablenotemark{b} & }
\tablecolumns{4}
\startdata
SGR 0501+4516 & 1.95 $\pm$ 0.04 & 2.0 & $3\times 10^{-4}$ \\
SGR 0526-66 & 50 & 7.3 & $6\times 10^{-7}$ \\
SGR 1627-41 &  11 $\pm$ 0.3 & 2.2 & $4\times 10^{-6}$ \\
SGR 1806-20 & 15.1$^{+1.8}_{-1.3}$ & 21 & $2\times 10^{-5}$ \\
SGR 1900+14 & 12--15 & 6.4  & $7\times 10^{-6}$ \\
1E 1547.0-5408\tablenotemark{c} & $\sim$ 9 & 2.2 & $5\times 10^{-6}$ \\
XTE J1810-197 & $\sim$ 5 & 1.7 & $1\times 10^{-5}$ \\
1E 1048.1-5937 & 2.7 $\pm$ 1 & 4.2 & $1\times 10^{-4}$ \\
AXP 2259+586 & 3.0 $\pm$ 0.5 & 0.59 & $1\times 10^{-5}$  \\
CXO J164710.2-455216 & $\sim$ 5 & 1.6 & $1\times 10^{-5}$ \\
\enddata 
\tablenotetext{$\dagger$}{Using Equation (\ref{elmillermod}) for 50
  keV photons.  Applying a GR correction would reduce this estimate by
$\approx 25$\%.}
\tablenotetext{a}{References for distances, in source order:
  \citet{Xu06}, \citet{Klose04}, \citet{Corbel99},
  \citet{Corbel04}, \citet{Vrba00}, \citet{Camilo07},
  \citet{Gotthelf04}, \citet{Gaensler05}, \citet{Kothes02},
  \citet{Clark05}.   Note that some of these distances are rather
  uncertain, which will affect the estimated
  critical flux.}
\tablenotetext{b}{References for inferred magnetic fields, in source order:
  \citet{Woods08}, \citet{Kulkarni03}, \citet{Esposito09},
  \citet{Mereghetti05}, \citet{Woods02}, \citet{Camilo07},
  \citet{Gotthelf05}, \citet{Gavriil04b}, \citet{Gavriil02},
  \citet{Israel07}. }
\tablenotetext{c}{Also designated as SGR J1550-5418.}
\end{deluxetable}

The last four sources listed in Table \ref{sdata} (XTE
J1810-197, 1E 1048.1-5937, AXP 2259+586 and CXO J164710.2-455216) have
never shown bursts with peak fluxes as high as the predicted critical values \citep{Gavriil02b,Gavriil04, Kaspi03, Woods05, Israel07}.  The other sources,
however, have shown brighter bursts that have reached or exceeded
the predicted critical flux.  

SGR 0526-66, SGR 1627+41, SGR 1900+14 and 1E 1547.0-5408 (SGR  
J1550-5418) have all had regular (short)
bursts with peak fluxes close to or exceeding the predicted limit
\citep{Golenetskii87, Gogus99, Woods99b, Woods99, Esposito08,
  Mereghetti09}.  A detailed study of those bursts that appear to
exceed the critical limit is beyond the scope of this paper.  For SGR
1806-20, however, which 
has a higher predicted critical luminosity due to its stronger field,
the smaller bursts do not reach the limit \citep{Gogus01}. 

The rarer intermediate bursts can also exceed the critical flux.  SGR
1900+14 had three intermediate bursts in 2001 that exceeded the predicted
critical flux \citep{Kouveliotou01,Ibrahim01}.  The events on April 28th and August
29th show no multipeak structure in their lightcurves.  However the
brightest event, on April 18th, does have an unusual feature:  at the
end of the outburst, the flux drops suddenly to near zero before there
is another peak \citep{Guidorzi04}. The lightcurve of this event is very similar to the
candidate PRE event from SGR 0501+4516.  One of the bursts discussed
by \citet{Israel08b}, on March 29th 2006, also reaches a peak flux of
$1\times 10^{-5}$ ergs/cm$^2$/s and 
looks like a multipeaked event.  At least one of the blackbody spectral
components that these authors fit expands and cools at the point where PRE
would occur if this was happening: \citet{Israel08b} comment on
the fact that the flux is close to critical, but do not discuss this
possible signature of PRE. 

All three giant flares exceeded the predicted limit by orders of
magnitude.  For SGR 0526-66 both initial flare and the detected
portion of the pulsating tail
exceed the limit \citep{Golenetskii87}.  For SGR 1900+14, peak flux in
the giant flare exceeded $3.4\times 10^{-3}$ ergs/cm$^2$/s
\citep{Hurley99}.  Flux would have dropped through the critical value
as the lightcurve decayed, but no odd behaviour is apparent in the
lightcurve at this time.  For SGR 1806-20, the
precursor to the giant flare reached a flux of $3.2\times 10^{-5}$
ergs/cm$^2$/s \citep{Boggs07}, very close to $F_\mathrm{crit}$.  The peak flux in
the main burst was far above the critical level.  The fluxes in the tail of the giant
flare do drop through the critical flux, and it is
interesting to note that the blackbody
component (there is an additional spectral component as well) remains
close to or lower than the critical flux \citep{Palmer05}.  It is
possible that this component has reached the magnetic Eddington limit.   

We note
that mass
ejection plays a significant role in giant flares, however, by
blowing scattering material away from the neutron star surface
\citep{Gaensler05b}.  \citet{Thompson95} noted that the amount of energy released in the
giant flares cannot be all trapped in the closed field region, and
proposed that the main initial part of the burst comes out along the
open field lines, driven by a wind from the pair plasma.  In this
case, PRE might not occur because the radiation pressure would be
converted to kinetic energy that would drive the atmosphere to escape
velocity. One extra condition for PRE to occur in magnetar bursts
might be that strong winds do not form and that the atmosphere remains
more or less static, or at least the motion does not reach escape
velocity.  In this case PRE in magnetars might occupy a fairly small
region of phase space between exceeding the magnetic Eddington limit
and driving a strong wind.

\section{Conclusions}
\label{conclusions}

We have examined the factors necessary for PRE to happen during
thermonuclear X-ray bursts,
and shown that they can also be met for some magnetar burst emission
models.  While additional conditions may also be necessary for PRE to
occur during magnetar bursts, the possibility certainly seems plausible.  An
unambiguous identification of PRE in a magnetar burst, however, will require better burst spectral
modelling taking into account the highly asymmetric emission and
scattering environment around the star. 

If magnetic PRE
can be identified conclusively then it could prove to be a very useful
tool.  It would constrain, for example,  the
trigger and emission mechanisms for magnetar bursts.
Identification of the magnetic Eddington limit also has
potential as a new constraint of the equation of state, provided that
the source distance and magnetic field strength can be measured by other means.   
We have argued that the August 24th 2008 burst from SGR 0501+4516
is a strong candidate for an open field line PRE burst.  However there
are other events that reach or exceed the predicted
critical fluxes in other magnetars.  To test the
consistency of our model these bursts must be given 
detailed consideration once better spectral and emission
models are in place.  

For magnetars PRE is certainly not inevitable
at a given flux, as it appears to be for X-ray bursts.  The occurrence
of PRE will depend on whether emission
occurs in open or closed field line regions, and on whether the
initial burst has been so strong that scattering material has been
blown away from the surface.  PRE may however help to explain some of
the extreme variability that we see in magnetar burst properties.

\acknowledgments

This publication is part of the GBM/Magnetar Key Project (NASA grant
NNH07ZDA001-GLAST, PI: C. Kouveliotou).  ALW acknowledges support from a Netherlands 
Organization for Scientific Research (NWO) Vidi 
Fellowship, and would like to thank Maxim
Lyutikov for discussions on magnetospheric triggers, and Duncan Galloway and Nevin Weinberg for discussions about
PRE in X-ray bursts.  AJvdH is supported by an appointment to the
NASA Postdoctoral Program at the MSFC, administered by Oak Ridge
Associated Universities through a contract with NASA.  EG and YK
acknowledge EU FP6 Transfer of Knowledge Project `Astrophysics of
Neutron Stars' (MTKD-CT-2006-042722).  MGB acknowledges support through
NASA grant NNX10AC59A and NSF grant AST-0607651.

\bibliography{magpre} \bibliographystyle{apj}

\end{document}